\documentclass[11pt,a4paper,dvipsnames]{amsart}
\usepackage{amsthm}
\usepackage{graphicx}
\usepackage{caption}
\usepackage{natbib}
\usepackage{subcaption}
\usepackage{amsfonts}
\usepackage{graphicx}
\usepackage{framed}
\usepackage{amssymb}
\usepackage{mathrsfs} 
\usepackage{tikz}
\usepackage{bm}

\usepackage{xcolor}
\definecolor{cornellred}{rgb}{0.7, 0.11, 0.11}

\usepackage[colorlinks=true,urlcolor=RoyalBlue,citecolor=RoyalBlue,linkcolor=RoyalBlue,linktocpage,pdfpagelabels,
bookmarksnumbered,bookmarksopen]{hyperref}

\usepackage{amsthm}
\usepackage{graphicx}
\usepackage{caption}
\usepackage{comment}
\usepackage{caption}

\usepackage{geometry}
\makeatletter
\@namedef{subjclassname@2020}{\textup{2020} Mathematics Subject Classification}
\makeatother

\numberwithin{equation}{section}
\title[Profiling vs. Case-specific Evidence]{Profiling vs. Case-specific Evidence: \\ A Probabilistic Analysis}

\begin{document}
\author[M.\,Di Bello]{Marcello Di Bello}
\author[N.\,Cangiotti]{Nicol\`o Cangiotti}
\author[M.\,Loi]{Michele Loi}

\address[M.\,Di Bello]{Arizona State University \newline\indent975 S. Myrtle Ave, Tempe, AZ 85287, United States} \email{mdibello@asu.edu}

\address[N.\,Cangiotti]{ Politecnico di Milano \newline\indent via Bonardi 9, Campus Leonardo, 20133, Milan, Italy}\email{nicolo.cangiotti@polimi.it}

\address[M.\,Loi]{Algorithmwatch\newline\indent Linienstra\ss{}e 13, 10178, Berlin, Germany}\email{michele.loi@icloud.com}


\begin{abstract}
The use of profiling evidence in criminal trials is a longstanding controversy in legal epistemology and evidence law theory. Many scholars, even when they oppose its use at trial, still assume that profiling evidence can be probative of guilt. We reject that assumption.
Profiling evidence may support a generic hypothesis, but is not
evidence that the defendant is guilty of the specific crime of which they are accused.
We contrast profiling evidence with case-specific evidence, which speaks more directly to the facts of the case. Our critique departs from others by grounding the argument in a probabilistic analysis of evidentiary value. We also explore the implications of our account for debates about stereotyping.
\end{abstract}
\maketitle

\section{Introduction}

The use of profiling evidence in criminal trials has long been a controversial topic in legal epistemology and evidence law theory. Consider a simple case. A burglary occurs, and police begin their search. Data show that repeat offenders commit a high proportion of burglaries in the area. Individuals with a prior burglary conviction are statistically far more likely to reoffend than those with no record. If a known burglar is found near the scene, does that make them more likely to be the perpetrator than someone without a criminal history? 
Does matching a criminogenic profile strengthen the case against a suspect? 

To be considered incriminating, profiling evidence must meet three conditions: (i) the defendant matches a  profile $P$ which is defined by stable characteristics such as age, upbringing, or criminal history; (ii) the defendant is  charged with a crime of type $X$, say first-degree murder, burglary, tax fraud; and (iii) empirical data show that individuals with profile $P$ commit crimes of type $X$ at a higher rate than the general population. To avoid gerrymandering, assume the correlation between profile $P$ and crime of type $X$ is causally plausible, and not a statistical coincidence or the result of arbitrary data grouping. If you walk down a sidewalk where everyone else deals drugs, matching the profile `on the drug dealer sidewalk' would not count as incriminating profiling evidence for the crime of drug dealing.\footnote{Well-known cases of naked statistical evidence---such as Prisoners in the Yard or Gatecrasher---would not count as examples of profiling evidence under this definition. The profile characteristics in those cases lack any non-incidental connection to the wrongdoing. Genuine profiling evidence rests on characteristics, such as prior convictions, demographic traits, or behavioral patterns, that are thought to be meaningfully linked to the type of offense at issue, rather than being incidentally correlated with it.}  

Most scholars agree that profiling evidence cannot, on its own, establish guilt beyond a reasonable doubt. But many also believe that it can bolster the case for guilt. Some go further and suggest it may be more probative of guilt than seemingly less controversial forms of evidence, such as eyewitness testimony  \citep{redmayne02}. Despite this, courts---particularly in the United States---often exclude profiling evidence \citep{KoehelerBaseRateRelevant}. For example, prior similar crimes are generally inadmissible to prove that the defendant committed the present offense. This resistance reflects a widely held intuition: individuals should not be judged by their past actions \citep{duff2021} or by traits they happen to share with others who commit crimes at higher rates \citep{colyvan01}.

If profiling evidence is genuinely probative of guilt, then this  intuitive resistance must either be mistaken or justifiable on other grounds. So a good bit of the literature has taken one of two routes: either that we should reconsider our resistance to profiling evidence \citep{Koehler1990Veridical-Verdi, redmayne02, picinali16}, or that other reasons make profiling evidence problematic. On moral or policy grounds, some argue that using profiling evidence undermines the  recognition of defendants as autonomous agents  \citep{Wasserman91, pundik2017};  others point out that verdicts based on profiling evidence weaken the deterrence function of  criminal trials  \citep{enoch_statistical_2012, dahlman2020}; others still contend that admitting such evidence unfairly increases the risk of false conviction for those who fit a criminogenic profile  \citep{di_bello_profile_2020}. But, differences aside, the literature often accepts one assumption: profiling evidence can be probative of guilt. We challenge this assumption. 

To argue that profiling evidence is not probative of guilt, we formalize an intuition that has been around for some time: profiling evidence is not case-specific  \citep{eggleston1980, thomsonIndEv1986, Stein05}. The idea of case-specific evidence has sometimes been dismissed as incoherent, on the grounds that no evidence can uniquely identify a perpetrator. We agree. All forms of evidence rely on generalizations and cannot be uniquely individualizing \citep{schauer_profiles_2009, tillers05}. But evidence need not be uniquely individualizing to be case-specific. On our account, evidence is case-specific if it helps to establish that the defendant committed a particular crime at a particular time and place. Profiling evidence does not meet this standard. At best, it helps to establish that the defendant committed a certain type of crime, not that they committed \textit{this} crime. 


Unlike others who have criticized profile evidence on epistemic grounds, we do not reject the view that legal evidence should be assessed probabilistically. Epistemic critiques allege that profiling evidence fails to promote certain non-probabilistic epistemic ideals, such as truth tracking \citep{enoch_statistical_2012, pritchard2015}, knowledge \citep{moss2018, Moss2023}, normic support \citep{smith2017}, or ruling out relevant alternatives \citep{Gardiner2019}. Our critique, by contrast, remains within a probabilistic account of probative value. For us, a piece of evidence is probative of a guilt hypothesis $G$ if considering the evidence should rationally increase one's degree of belief (probability) in 
$G$. 
The crux of our argument is that profiling evidence does not raise the probability of the \textit{specific} guilt hypothesis and thus fails to be probative of guilt.

The plan is as follows. We start with an argument why profiling evidence can be  probative of guilt (\S \ref{sec:seems-probative}), and explain why it is flawed (\S \ref{sec:not-probative}). Next, we contrast profiling  and case-specific evidence (\S \ref{sec:diagnostic}) and address the objection that generalizations are ubiquitous in legal reasoning (\S \ref{sec:generalizations}). Finally, we show that specificity is implicit in epistemic critiques of profiling evidence (\S \ref{sec:other-espistemic}). In the conclusion, we  explore the implications of our account
for debates about stereotyping (\S \ref{sec:stereotyping}). 


\section{Profiling evidence seems probative}
\label{sec:seems-probative}






Recall the burglary example from the beginning. Suppose a suspect is found near the crime scene and fits the profile of a prior offender. Should this information increase the probability that they committed the crime? To answer this, we need a principled way to update our degrees of belief (probabilities) based on new information. Bayes' theorem provides a natural framework for this purpose. The odds form of the theorem is  useful  for simplifying calculations:

\begin{equation*}
    \frac{\mathbb{P}(G \mid E)}{\mathbb{P}(I \mid E)} = \frac{\mathbb{P}(E\mid G)}{\mathbb{P}(E \mid I)} \times \frac{\mathbb{P}(G)}{\mathbb{P}(I)}\,.
\end{equation*}

\noindent
The prior odds $\frac{\mathbb{P}(G)}{\mathbb{P}(I)}$
represent our initial belief about the likelihood of guilt $G$ versus innocence $I$  before considering the new evidence. When we obtain new evidence $E$, we update these odds by multiplying by the likelihood ratio $\frac{\mathbb{P}(E\mid G)}{\mathbb{P}(E \mid I)}$
 which measures how much more (or less) likely the evidence is if the suspect is guilty rather than innocent. This results in the posterior odds $\frac{\mathbb{P}(G \mid E)}{\mathbb{P}(I \mid E)}$ which reflect our revised belief after accounting for the evidence.

In our example, the evidence 
 is that the suspect fits the profile of a prior offender, and the guilt hypothesis is that they committed the burglary. Our goal is to determine how much this evidence should change our belief about the suspect’s guilt.
%
%
The prior odds 
\[
\frac{\mathbb{P}(\textit{burglary})}{\mathbb{P}(\textit{NOT burglary})}
\]
will increase so long as the likelihood ratio 
\[
\frac{\mathbb{P}(\textit{prior offender} \mid \textit{burglary})}{\mathbb{P}(\textit{prior offender} \mid \textit{NOT burglary})}
\]
is  greater than one. Suppose data show that 80\% of the burglaries in the area are committed by prior offenders, so the numerator
$\mathbb{P}(\textit{prior offender} \mid \textit{burglary})$ equals $80\%$.
The denominator can be set to the share of prior offenders in the population, presumably a low percentage, say $1\%$.\footnote{By the law of total probability, $\mathbb{P}(\textit{prior offender})$  equals 
\[\mathbb{P}(\textit{prior offender} \mid \textit{burglary})\mathbb{P}(\textit{burglary}) + \mathbb{P}(\textit{prior offender} \mid \textit{NOT burglary})\mathbb{P}(\textit{NOT burglary})
\]
 Therefore, by algebra, we have:
\[
\mathbb{P}(\textit{prior offender} \mid \text{NOT burglary}) 
= \frac{
\mathbb{P}(\textit{prior offender})}
{\mathbb{P}(\textit{NOT burglary})} 
- 
\frac{\mathbb{P}(\textit{prior offender} \mid \textit{burglary}) \mathbb{P}(\textit{burglary})}{{\mathbb{P}(\textit{NOT burglary})}}\,.
\]
Since $\mathbb{P}(\textit{burglary})$ is low and $\mathbb{P}(\textit{NOT burglary})$ high, the denominator goes to one and the second term goes to zero. So $\mathbb{P}(\textit{prior offender} \mid \textit{NOT burglary}) 
\approx \mathbb{P}(\textit{prior offender})$.}
The likelihood ratio, then, will be roughly equal to 80. 
In other words, the priors odds of guilt should increase by a factor of 80 after taking into account  that the defendant is a prior offender. The profiling evidence seems strongly probative of guilt.

For a real case, consider  United States v.\ Vue (13 F.3d 1206, 8th Circuit, 1994) in which two Hmong brothers were accused of drug-related offenses in Minneapolis. 
The data available to law enforcement showed that 95\% of the smuggling cases in  Minneapolis  involved people of Hmong ancestry, while the Hmong were 6\% of the population.
%
%
By the reasoning above, the likelihood ratio 
\[
\frac{\mathbb{P}(\textit{Hmong } \mid \textit{drug trafficking})}{\mathbb{P}(\textit{Hmong} \mid \textit{NOT drug trafficking})} \approx \frac{\mathbb{P}(\textit{Hmong} \mid \textit{drug trafficking})}{\mathbb{P}(\textit{Hmong})}
\]
should equal 
\[
\frac{95\%}{6\%}\approx 15\,.
\]
The prior odds of committing drug trafficking must increase 15 times upon learning that the suspect fits the profile `Hmong'.  
This again shows that fitting the profile `Hmong' can be strongly probative of guilt in a drug  case in Minneapolis. 


How high will the posterior guilt probability be after taking profile evidence into account? In the Vue case, there was also circumstantial evidence against the brothers: a package containing drugs was sent to their address. Suppose this evidence alone raises their probability of guilt to 50\%, so the prior odds are 1:1. Since they also match the profile `Hmong', their probability of guilt would jump to more than 90\%.\footnote{Given a likelihood ratio of 15, the posterior odds would be 15:1 that they committed drug trafficking, so their posterior probability of guilt would be $15/(15+1)\approx 93\%$.}
 The same holds in the burglary case. Suppose additional circumstantial evidence is found, such as that the suspect’s shoe size matches the footprints at the crime scene.  If we combine this with profiling evidence of prior burglary, the probability of guilt could approach 100\%.\footnote{If the likelihood ratio is 80 and the prior odds are (conservatively) 1:1  based on the footprints, the posterior odds would be 80:1, with a posterior guilt probability of $80/(80+1)\approx 99\%$.}
 

Still, there is something unsettling about concluding that a defendant is almost certainly guilty---let alone convicting them---when a significant part of that conclusion is based on profiling evidence, such as a prior burglary record or demographic factors like Hmong ancestry in Minneapolis.
Profiling evidence seems to play an outsized role in the argument for guilt. Should we trust this intuition, or is it the result of cognitive bias and a lack of familiarity with probabilistic reasoning? Is our discomfort a sign of a genuine flaw in the reasoning?

\section{Profiling evidence isn't probative}
\label{sec:not-probative}

The intuitive resistance to using profiling evidence in trials is not simply due to a misunderstanding of probabilistic reasoning. In fact, we think this uneasiness can be explained in probabilistic terms. It is customary to note that profiling evidence is based on general statistical correlations---broad patterns rather than case-specific facts. Suppose that since the suspect fits a certain profile, they are statistically more likely to  commit a burglary or drug offense at some point in their life. Even if this general claim is true, it does not follow that the suspect is more likely to have committed this particular burglary or drug crime. 

We should not conflate two different hypotheses. The \textit{generic} hypothesis that the suspect committed (or will commit) an offense of this type at some point differs from the \textit{specific} hypothesis that the suspect committed (or will commit) this particular offense.
Since criminal trials are about determining responsibility for specific crimes, relying on profiling evidence leads to an unjustified leap in reasoning. Profiling evidence may support a  generic hypothesis, but is not  evidence that the defendant is guilty of the specific crime currently on trial.

Our argument against profiling evidence can be broken down as follows:

\begin{itemize}

\item[(P1)] The hypothesis of guilt in a criminal trial must be sufficiently specific, typically including details about when and where the crime occurred, the modus operandi, and the identity of the perpetrator.
 
\item[(P2)] Evidence is probative of guilt only if it increases the probability of the (specific) guilt hypothesis that the prosecution must establish.

\item[(P3)] Profiling evidence does not  increase the probability of the (specific) guilt hypothesis that the prosecution must establish (or does so minimally). 

\item[\textit{So:}]  Profiling evidence is not probative of guilt.
\end{itemize}

\noindent
The first premise applies to most criminal offenses---for example, murder, theft, drug trafficking---where establishing who committed a specific crime at a specific time and place is essential. Some crimes may not require such specificity, but this does not undermine the general point. The second premise is more controversial, but it is a concession  to our opponents who  rely on probability theory to argue that profiling evidence can be probative of guilt. Our main focus is on defending the third premise which rests on the distinction between generic and specific.

Consider the generic standpoint. Let $G_g$ stand for the generic guilt hypothesis that a person committed (or will commit) an offense of a certain type, say a drug offense. Let $P$ be the proposition that the defendant fits an incriminating profile, like being a prior offender. By the argument in the previous section, the likelihood ratio $\frac{\mathbb{P}(P \mid G_g)}{\mathbb{P}(P \mid NOT \ G_g)}$ can be several times greater than one. So profiling evidence is indeed strongly probative of the generic hypothesis $G_g$ that they committed (or will commit) an offense somewhere at some point. This conclusion, however, does not tell us whether  profiling evidence can be probative of the hypothesis that the suspect committed \textit{this} drug crime for which they are charged.  More formally, even if the generic likelihood ratio $\frac{\mathbb{P}(P \mid G_g)}{\mathbb{P}(P \mid NOT \ G_g)}$ is many times greater than one, it does not follow that the specific likelihood ratio $\frac{\mathbb{P}(P \mid G_s)}{\mathbb{P}(P \mid NOT \ G_s)}$ must be as well,
where $G_s$ stands for the hypothesis that the suspect is guilty of this crime. 

%
%


Let us look at the specific likelihood ratio more closely. Start with the denominator $\mathbb{P}(P\mid  NOT \  G_s)$, the probability that the suspect fits the profile (say, being Hmong) assuming they did not commit this specific crime. This probability can be equated to the profile's population rate:
\[\mathbb{P}(P \mid NOT \ G_s)\approx \mathbb{P}(P \mid NOT \ G_g) \approx \mathbb{P}(P)\,.
\]
The reason for this approximation is that if crimes of this type are rare, most people who didn’t commit this specific crime also didn’t commit any such  crime at all.\footnote{Let $G_s$ be the hypothesis that the defendant committed \emph{this} crime; $G_g$ the hypothesis that they committed \emph{some} crime of this type; and $P$ the profile. For anyone innocent of this specific crime ($ NOT \  G_s$), distinguish those who never committed a crime of this type ($ NOT \  G_g$) from those who committed one but not this one ($G_g \ \text{and}\ NOT \  G_s$). By the law of total probability:
\[
\begin{split}
\mathbb{P}(P \mid  NOT \  G_s)
=
\mathbb{P}(P\mid  NOT \  G_g)\, \mathbb{P}( NOT \  G_g\mid  NOT \  G_s)
\, +\\
+\, \mathbb{P}(P\mid G_g ,  NOT \  G_s)\, \mathbb{P}(G_g ,  NOT \  G_s\mid  NOT \  G_s)\,.
\end{split}
\]
Since crimes of this type are rare, the second class is extremely small, so the second term is negligible. For the much larger first class, the prevalence of the profile is simply its population base rate: $\mathbb{P}(P\mid  NOT \  G_g)\approx \mathbb{P}(P)$. Hence $\mathbb{P}(P\mid  NOT \  G_s)\approx \mathbb{P}(P)$.} So the denominator is easy. What about the numerator $\mathbb{P}(P\mid  G_s)$, the probability that the suspect fits the profile  given that they did commit this offense? We will see that the numerator cannot be assigned a precise value.  

At first, it is tempting to run a  quick argument.  Since the crime under consideration is a specific event in time and place, whereas the profile represents a stable, long-term characteristic of a person, we might think that the occurrence of a particular crime would not affect at all whether the suspect fits the profile. The two belong to different temporal scales: the crime is episodic, while the profile captures enduring features such as ethnicity, gender, socio-economic status, etc. For that reason, it would seem plausible to equate the numerator to the  prevalence of the profile in the population:
\[\mathbb{P}(P \mid G_s)\approx \mathbb{P}(P)\,.
\]
Under this line of reasoning, the specific likelihood ratio would equal one, and the profiling evidence would be irrelevant to the question of specific guilt.


One could reply that, in absence of any more specific information, we should rely on the information we have. In other words, the probability that someone would have the profile given that they committed this specific crime should be equated to the probability that someone  would have the profile given that they committed a crime of the type in question:
\[\mathbb{P}(P \mid G_s)\approx \mathbb{P}(P \mid G_g)\,.
\]
This approximation assumes that the mechanism by which individuals acquire the profile remain invariant across different refinements of the population of offenders, from generic to more specific. If this is right,  then the specific and general likelihood ratios would behave the same. If general likelihood ratio is high, the specific likelihood ratio must be high as well. Profiling evidence would be probative of specific guilt as much as it would be of generic guilt. 

So, which one is it? Is profile evidence strongly probative or irrelevant for establishing the specific guilt hypothesis? We think it is neither. 
%
%
When we reason about generic guilt $G_g$---that someone committed \emph{some} crime of type $X$ at \emph{some} time and place---we can rely on the overall distribution: 80\% of type-$X$ offenders have profile $P$. 
But when the hypothesis becomes \emph{particularized}---that the suspect committed this specific $X$-crime---the relevant probabilities can shift dramatically.  Specific crimes differ from one another by location, time of day, method, etc. These characteristics will change the demographic mix of the population of offenders. 
For instance, the subclass of burglaries that occurred in a wealthy neighborhood during daytime, using sophisticated lock-picking techniques, etc. may have a very different distribution of the profile $P$ in question than the class of burglaries as a whole. 

The invariance assumption---that the conditional probability remains stable as we move from generic to specific hypotheses---is unwarranted. 
  Unless we have grounds to believe that the specific crime under consideration was generated by a mechanism representative of the overall class of burglaries, we cannot responsibly assume $\mathbb{P}(P \mid G_s)$ equals the generic value $\mathbb{P}(P \mid G_g)$. Nor can we assume that it equals the base rate $\mathbb{P}(P)$. We simply have no information to make these approximations. 
The numerator $P(P \mid G_s)$ is effectively unknown. 
With an unknown numerator, the specific likelihood ratio 
\[
\frac{P(P \mid G_s)}{P(P \mid \text{NOT } G_s)}\]
cannot be assessed either. For a more detailed argument, see the \nameref{Appendix}.

Consider an analogy with eyewitness identification. If we lack information about the specific conditions under which an eyewitness made an identification---such as distance, lighting, duration of exposure, stress level, the presence of a weapon, or whether the witness was wearing corrective lenses---then we have no reliable basis for assessing the accuracy of that identification. Without knowing which of those factors applied, we cannot meaningfully estimate the probability that the identification was correct. For all we know, the eyewitness may have been observing under conditions so poor that their judgment was no better than chance. This means that the probative value of the eyewitness testimony is undefined: it might be high, if the conditions were optimal, or negligible, if they were not. The proper response to missing information about identification conditions is not to average over all possible cases, but to acknowledge that we cannot tell whether the testimony provides evidence at all. 

If our argument is right, we should expect less resistance to profiling evidence when a lower degree of specificity is required. Consider, for example, pretrial decisions about reoffending. In these contexts, the goal is not to determine whether someone committed a specific crime at a specific time and place. Decisions in the pretrial context determine whether a defendant should be released, required to post bail, or detained until trial. 
To make these assessments, judges may consider factors like prior criminal history, age, employment status, and substance use history. 
Here, profiling evidence is  generally seen as less problematic. We conjecture that part of the story is that, in pretrial decisions about reoffending, judges assess the likelihood of any future offense, rather than determining responsibility for a particular future crime.\footnote{Racial profiling is banned for ethical and legal reasons, but other forms of profiling---provided they do not collapse into racial profiling---are not prohibited in pretrial risk assessments.} The distinction generic/specific highlights why profiling evidence is  controversial in criminal trials, where specificity is essential, but more accepted in settings focused on assessing general risk.

\section{Case-specific evidence}

\label{sec:diagnostic}

It is helpful to contrast profiling evidence with \textit{case-specific evidence}, which is more clearly tied to a particular act. 
Evidence is case-specific if it consists of spatio-temporally specific facts, for example, that a young men was seen around the crime scene at the relevant time. 
Profiling evidence does not typically include such facts. Learning that young men tend to commit crimes of type $X$ more often than older men---and the defendant is a young man tried for a crime $X$---makes no reference to specific spatio-temporal facts.  

The probative value of case-specific evidence can be assessed with likelihood ratios. Suppose that, upon searching John's suitcase, authorities find packets of illegal drugs with his fingerprints on the packaging ($F$). This evidence concerns his involvement in this specific crime ($G_s$). The likelihood ratio  will be as follows:
\begin{equation*}
    \frac{\mathbb{P}(F \mid G_s)}{\mathbb{P}(F\mid NOT \ G_s)}\,.
\end{equation*}
Here, $\mathbb{P}(F \mid G_s)$---the probability of finding John’s fingerprints on the drug packets given that he committed this trafficking act---is high, say like before 80\%. Because such evidence is not often found on innocent people, $\mathbb{P}(F\mid NOT \ G_s)$ must be low. Given this evidence, the probability of specific guilt should increase significantly. Unlike with profiling evidence, what increases here is not just the probability of John committing a drug-related crime, but also---and most importantly---the probability of John committing this specific drug-related crime.

Case-specific evidence is probative of the specific guilt hypothesis because of a  presumed causal connection between 
the facts that constitute the crime and the facts that constitute the evidence. Both facts
must be spatio-temporally specific and cohere with one another, for example, a burglary that occurred at 10:42 PM on a particular street, and an eyewitness testimony that a young man was seen fleeing the scene of the burglary after it occurred. In evaluating the evidence, we implicitly construct a causal model---also spatio-temporally structured---that links the defendant’s presence to the crime. The causal connection may later be proven wrong, but  it is initially presumed until other evidence shows otherwise. 

What happens when case-specific evidence links the suspect to the crime and profiling evidence is used to corroborate this evidence? 
Some might reason that, when case-specific and profiling evidence are combined, profiling evidence  increases the probability of the generic guilt hypothesis as well as the specific one. We disagree. Our claim that profiling evidence only increases the probability of the generic guilt hypothesis holds regardless. 

In the case of the Vue brothers, a package containing drugs was delivered to their address. That is an example of case-specific evidence, although of a weak kind. It indicates---fallibly of course---they were linked to the specific act of drug trafficking. Now, once we learn that they were Hmong and that they are more likely to commit some drug trafficking crime in Minneapolis, compared to people in other ethnic groups, what are we to conclude? We have very weak additional evidence that they committed the specific act of drug trafficking. More schematically, 
suppose fallible, case-specific evidence $E1$ supports the claim that the defendant committed this specific burglary, so $\mathbb{P}(G_s \mid E1)\approx 50\%$. In addition, suppose infallible evidence $E2$ establishes that the defendant committed (or will commit) a burglary at some point somewhere, so $\mathbb{P}(G_g \mid E2)\approx 100\%$. Does the combination of $E1$ and $E2$ raise the probability of the specific hypothesis $G_s$? It is not clear how. We would need additional information to connect the two. 

To conclude this section, we consider a typology of evidence that is closely related to, but different from, case-specific evidence, namely diagnostic evidence, also often called trace evidence. It is tempting to conflate two distinctions: profiling v. case-specific evidence and predictive  v. diagnostic or trace evidence. 
Relative to the facts that constitute the crime, diagnostic or trace evidence is causally downstream and backward-looking: it is evidence that the crime occurred \citep{mackor21, Lee-Stronach2023}. Predictive evidence, by contrast, is forward-looking: it raises the likelihood that an individual will commit an act in the future. At first, profiling appears problematic because it is predictive rather than diagnostic. Since judgments of criminal liability are backward-looking, profiling evidence \textit{as} predictive seems unfit to justify these judgments. 

But this way of framing the issue  runs into a difficulty. Consider motive evidence: it is predictive, but it is perfectly acceptable evidence. The crux is not whether the evidence is predictive or diagnostic, but whether or not it is specific to the crime charged. Predictive evidence can, in principle, be case-specific.  Motive evidence is predictive, in that having a motive increases the likelihood that someone would commit a crime at a later time. It is also case-specific as it can be tied to the facts of the case, for example, if the defendant had a documented conflict with the victim. Admittedly, it is harder to think of diagnostic evidence that is not case-specific. Which explains why diagnostic (or trace) evidence and case-specific evidence tend to be conflated.

\section{The ubiquity of generalizations}

\label{sec:generalizations}

Our account of the epistemic deficiency of profiling evidence rests on a simple   point: profiling evidence does not support the specific guilt hypothesis. It may increase the probability that the defendant committed or will commit some crime of a certain type, but it does not increase the probability that the defendant committed this crime.   One might be tempted to go further and say this: profiling evidence merely shows that someone like the defendant---not specifically \textit{this} defendant---is statistically more likely to commit a certain type of crime. This formulation, however, overstates the problem. It implies two flaws in profiling evidence: first, a lack of spatio-temporal specificity; second, an impermissible group-to-individual inference. The first  flaw is correct. The second incorrectly  assumes that evidential reasoning must  uniquely identify an individual as the perpetrator. But no piece of evidence uniquely identifies a person in isolation. All evidential reasoning relies on general categories. Even the most discriminating forms of evidence---say, a DNA match---only point to an individual by way of identifying traits that could be shared by multiple individuals.

The same, non uniquely identifying feature could be used in profiling or case-specific evidence. Suppose an eyewitness has seen a young man at the crime scene, and  the defendant is a young man. Thus, the witness testimony makes him a more likely perpetrator. Now consider this. The defendant is a young man accused of crime of type $X$, and data show that young people are more likely than older people to commit crimes of type $X$. Thus, the defendant is more likely to be the perpetrator once we learn they are young. The first inference is unobjectionable; the second is objectionable. But in both cases we have a non-uniquely identifying feature, being a young man. 
%
Anyone on trial who matches the characteristic 
`young man' would be incriminated by either form of evidence, far from a unique individualization.\footnote{The contrast can be made even starker by replacing `young man' with `Black man'. That a witness saw a Black man near the crime scene warrants the inference that the defendant---if he is a Black man---is more likely to be the perpetrator, but no demographic profiling evidence would warrant such an inference \citep{Banks2001Race-Bases-Susp}.} 
What is going on here?


The key is move away from the requirement of unique individualization and turn instead to that of spatio-temporal specificity. Specificity requires that the evidence and the crime are both anchored in particular times and places and that there is a presumed causal connection between them. A piece of evidence is spatio-temporally specific if it helps to construct or support a causal narrative that plausibly links the defendant to this crime at this time and this place. That is what gives case-specific evidence its probative force—not that it isolates the defendant as the only possible culprit, but that it is part of a coherent, causally grounded account of how the crime occurred.

\section{Epistemic accounts}
\label{sec:other-espistemic}

We will now contrast our account of why profile evidence is epistemically deficient ---it does not support the specific guilt hypothesis---with other epistemic accounts in the literature.  Profiling evidence has been criticized for failing to meet other epistemic standards, such as truth tracking, normic support, ruling out relevant alternatives or contributing to knowledge. We will argue that these accounts all implicitly assume a requirement of specificity. 

Truth tracking requires that evidence co-varies with the facts. 
For example, a fingerprint at the crime scene or a witness identifying the defendant tends to satisfy this requirement. If the defendant is guilty, we expect the fingerprint or identification to be there; if they are innocent, we do not. 
Profiling evidence, by contrast, fails to co-vary with the facts in this way. Whether or not the defendant committed the crime, the profiling evidence would remain unchanged. Since it depends on general demographic patterns, not the  facts of the case, profiling evidence would likely still be present regardless \citep{enoch_statistical_2012}.

But this raises a puzzle. To state the puzzle, first note that no piece of evidence perfectly tracks the truth. So tracking must be a matter of degrees. In probabilistic terms, evidence $E$ is truth tracking in a minimal sense if its likelihood ratio is greater than one \citep{Roush2006}:
\[\frac{\mathbb{P}(E \mid G)}{\mathbb{P}(E \mid I)}>1\,,
\]
where 
$G$ is the guilt hypothesis and 
$I$ is the innocence hypothesis. As seen earlier, this means the evidence is more likely to occur if the defendant is guilty than if they are innocent. 
If this is right, we have a puzzle. For one thing, if the likelihood ratio is greater than one, profiling evidence is truth tracking. For another, since it would be there whether the defendant is guilty or innocent, profiling evidence is not truth tracking. Both claims seem plausible. This is the puzzle. 

The answer lies in distinguishing generic and  specific hypotheses. The likelihood ratio may be greater than one for a generic hypothesis, for instance, that the defendant committed or will commit a crime of this kind. But it is not greater than one for the specific hypothesis that the defendant committed this particular crime. Evaluated  relative to the specific hypothesis, profiling evidence does not discriminate between guilt and innocence. The likelihood of seeing the profiling evidence is roughly the same whether the defendant is guilty or not. Which makes the likelihood ratio close to one. Profiling evidence can be truth tracking for broad, general hypotheses but fails to track the truth when applied to specific crimes. This explains why it feels both probative in some sense and epistemically deficient in another. So specificity resolves the apparent paradox.  

Normic support offers a different but related standard for epistemic justification. Roughly, evidence normically supports a proposition when, if the proposition were false, the presence of the evidence would be an anomaly calling for an explanation. Case-specific evidence typically satisfies this condition. For instance, finding the defendant’s fingerprints at the crime scene, or recovering stolen property in their possession, would be surprising if they were innocent. Their innocence would require an explanation: fabrication, contamination, or coincidence. Profiling evidence, by contrast, fails to meet normic support.  Innocent people matching a criminogenic profile are common, even expected. If the defendant were innocent, there would be nothing abnormal about the presence of the profiling evidence \citep{smith2017}. 

But the notion of normic support is elusive.  Why is it, exactly, that case-specific evidence would call for an explanation but profiling evidence would not, should the defendant turn out to be innocent? The answer lies in specificity. 

With case-specific evidence, we expect that the evidence is connected to the crime through a specific causal process. This process is concrete, occurring at a particular time and place. If someone’s fingerprints are found on a weapon, we assume they touched it during the crime. If it turns out they did not commit the crime after all, we naturally want an explanation. How did the fingerprints still get there? The falsity of the hypothesis that the person committed the crime triggers a need for an explanation because we expected a causal link between the evidence and the event. By contrast, profiling evidence does not enter in a specific, identifiable causal process originating from the particular facts of the crime. If someone who fits a profile turns out to be innocent, the  need for an explanation does not arise. There is no broken causal chain to investigate, because none was assumed in the first place. Once again, the key difference is specificity. Case-specific evidence suggests a particular cause-and-effect chain that invites explanation if it fails. Profiling evidence does not.

Finally, consider the idea of relevant alternatives. One critique of profiling evidence is that it doesn’t rule out all---or at least some of---the relevant alternatives to the guilt hypothesis, for example, the possibility that the defendant was elsewhere at the time of the crime, or that someone else with a similar profile committed it \citep{Gardiner2019, moss2018, Moss2023}. Once again, we think the crux of this account implicitly assumes a requirement of specificity.\footnote{We do not discuss the knowledge account explicitly, but note that eliminating relevant alternatives to $G$ is one way of gaining knowledge of $G$.}   

No evidence can ever rule out a relevant alternative completely. Ruling out alternatives is always a matter of degrees. Evidence makes some hypotheses more likely and others less likely. For example, if the evidence makes it more likely that the defendant was in Location A, then it becomes less likely they were in Location B. Rarely does the probability drop to zero. So the question is, which alternatives need to be ruled out---or made sufficiently less likely---for the evidence to be probative? The answer is: the alternatives that compete with the specific guilt hypothesis about the crime, grounded in particular details like time, place, and modus operandi. In other words, what matters are the spatio-temporally specific alternative hypotheses.

The point of the relevant alternatives account can be stated probabilistically as follows: profiling evidence lacks the specificity needed to make particular alternatives less likely. This claim follows from our earlier claim that  profiling evidence does not increase the probability of the specific guilt hypothesis. If profiling evidence does not increases the probability of the specific guilt hypothesis, then it will not decrease the probability of specific alternative hypotheses either.

\section{Conclusion}
\label{sec:stereotyping}
\label{sec:conclusion}




We have argued that profiling evidence, while it may appear to support a finding of guilt, is not probative of the defendant's guilt. Such evidence may increase the probability that the defendant committed (or will commit) a certain type of crime, but it does not increase the probability that they committed the crime for which they are on trial. 

Unlike other accounts in the literature, our argument does not rely on moral or policy considerations, nor does it require departing from a probabilistic assessment of the evidence. Under the Federal Rules of Evidence, evidence is considered probative only if (a) it increases the probability of a hypothesis, where (b) that hypothesis is relevant to deciding the case---that is, it is `of consequence in determining the action.' Profiling evidence satisfies the first condition: it raises the probability of a general guilt hypothesis. But it fails to satisfy the second condition because generic guilt is usually not at issue in a criminal trial.

We conclude by gesturing toward an application of our argument beyond the criminal trial to the topic of stereotyping. To illustrate, consider the following two scenarios. In Scenario A, an individual is walking from a wealthy, predominantly white neighborhood into another area typically associated with a specific ethnic profile and known for higher crime rates. The individual becomes more alert, as they fear a greater risk of being robbed. In this case, the individual is tracking a statistical correlation between place, demographic characteristics, and crime, in a context where no case-specific  accusation is made. Scenario B presents a different dynamic. The same individual has just been robbed. Looking around, they see two people:  Remus with ethnic attributes  associated with high-crime areas, and Dedus with a background typically linked to safer neighborhoods. The individual, without further evidence, asks Dedus to help restrain and search Remus.

We take no position on whether either action is justified on moral or political grounds. Our focus is epistemic. In Scenario A, the action is a precaution, not an attribution of culpability. In Scenario B, however, the individual makes a judgment of \textit{prima facie} culpability based on stereotyping---a case of statistical resentment \citep{enoch2021}. That judgment of culpability is epistemically unwarranted: there is no case-specific evidence linking Remus to this robbery and thus no evidential support for the specific hypothesis that Remus is culpable. 
%
The probability $\mathbb{P}(H_g \mid P)$ may be high while $\mathbb{P}(H_s \mid P)$ remains low, with the inferential chain weakening in the transition from generic  $H_g$ to specific  $H_s$.\footnote{Our argument does not depend on rejecting the empirical claim that stereotypes sometimes reflect statistical regularities at the group level. Even if we grant the claim that some stereotypes contain ``kernels of truth'' in terms of correlational accuracy \citep{jussim2012social}, the  problem remains of inferring a specific hypothesis from group statistics.} Like profiling evidence, stereotyping is abused when one moves carelessly from statistical patterns observed at the group level to judgments about a  spatio-temporally specific hypothesis. 

\section*[Appendix]{Appendix: A Multi-level Nesting Model}
\label{Appendix}

In this appendix, we provide a brief formal treatment of the \emph{Multi-level Nesting Model} (MNM), which we introduced in Section \ref{sec:not-probative} to explain why the invariance assumption $\mathbb{P}(P \mid G_s) \approx \mathbb{P}(P \mid G_g)$ is generally unwarranted.

Consider the population of all individuals who commit crimes of type $X$, such as burglaries or drug trafficking offenses. Let $P$ denote the property of possessing profile $P$ (e.g., being a prior offender or belonging to a certain demographic group), and let $G_g$ denote the generic hypothesis that an individual committed a crime of type $X$ at some unspecified time and place. The key insight is that this population is not homogeneous. Rather, it can be partitioned into $n$ distinct subpopulations or \emph{contexts} $S_1, S_2, \ldots, S_n$, where each $S_i$ corresponds to a specific set of circumstances under which crimes of type $X$ occur. For instance, in the case of burglaries, we might have: $S_1$ representing crimes in wealthy neighborhoods during daytime; $S_2$ representing crimes in low-income neighborhoods at night; $S_3$ representing crimes involving sophisticated lock-picking methods; and $S_4$ representing opportunistic crimes with simple methods. Each subpopulation $S_i$ may have its own prevalence of profile $P$ among offenders, which we denote $\mathbb{P}(P \mid G_g, S_i)$. Crucially, these prevalences need not be equal across subpopulations.

When we assess the generic guilt hypothesis $G_g$, we make no commitment to which subpopulation is relevant. By the law of total probability, the overall prevalence of profile $P$ among type-$X$ offenders is given by:
\[
\mathbb{P}(P \mid G_g) = \sum_{i=1}^{n} \mathbb{P}(P \mid G_g, S_i) \times \mathbb{P}(S_i \mid G_g)\,.
\]
This is a weighted average of the prevalences across all subpopulations, where the weights $\mathbb{P}(S_i \mid G_g)$ represent the proportion of type-$X$ crimes that occur in context $S_i$. This weighted average is precisely what empirical criminological data typically provide: aggregate statistics that lump together crimes occurring under diverse circumstances.

Now consider a specific guilt hypothesis $G_s$: the defendant committed \emph{this particular} crime of type $X$ at a specific time, place, and manner. Case-specific evidence---such as the location, time of day, modus operandi, and victim characteristics---typically narrows down which subpopulation(s) are relevant to the specific crime under consideration. In many cases, this evidence effectively identifies a single subpopulation $S_k$. Under these circumstances, the probability that the defendant has profile $P$ given that they committed this specific crime is:
\[
\mathbb{P}(P \mid G_s) = \mathbb{P}(P \mid G_g , S_k)
\]

The invariance assumption would assert that $\mathbb{P}(P \mid G_s) \approx \mathbb{P}(P \mid G_g)$, which would require that:
\[
\mathbb{P}(P \mid G_g, S_k) \approx \sum_{i=1}^{n} \mathbb{P}(P \mid G_g, S_i) \times \mathbb{P}(S_i \mid G_g)
\]
This equality can hold only if one of two conditions is satisfied. First, the \emph{uniformity condition}: the prevalence of profile $P$ is approximately the same across all contexts, i.e., $\mathbb{P}(P \mid G_g, S_i) \approx \mathbb{P}(P \mid G_g, S_j)$ for all $i, j$. Second, the \emph{representativeness condition}: the specific context $S_k$ happens to be ``typical'' in the sense that $\mathbb{P}(P \mid G_g, S_k)$ equals (or is very close to) the weighted average across all contexts. Neither condition can be assumed without substantial empirical evidence. Against uniformity, there is no reason to believe that the distribution of profile $P$ remains constant across contexts that vary significantly in location, time, method, victim demographics, and other factors. Indeed, criminological research consistently reveals substantial heterogeneity in offender characteristics across different crime contexts. Against representativeness, without detailed knowledge of how $P$ is distributed across the various subpopulations, we have no basis for assuming that the particular context $S_k$ identified by case-specific evidence is representative of the overall average. The invariance assumption is therefore epistemically unwarranted.

To illustrate this point concretely, suppose there are four subpopulations of burglars, each accounting for 25\% of all burglaries, so that $\mathbb{P}(S_i \mid G_g) = 0.25$ for all $i = 1, 2, 3, 4$. Suppose further that the prevalence of prior offenders varies substantially across these subpopulations: in wealthy areas during daytime ($S_1$), only 40\% of offenders have prior convictions; in wealthy areas at nighttime ($S_2$), this rises to 70\%; in low-income areas during daytime ($S_3$), it is 90\%; and in low-income areas at nighttime ($S_4$), it reaches 95\%. The generic probability is then:
\[
\mathbb{P}(P \mid G_g) = 0.25 \times 0.40 + 0.25 \times 0.70 + 0.25 \times 0.90 + 0.25 \times 0.95 = 0.7375 \approx 74\%\,.
\]
Suppose now that case-specific evidence---the location and time of the crime---indicates that the crime occurred in context $S_1$ (wealthy area, daytime). Then the probability that the perpetrator has a prior conviction, conditional on the specific crime, is:
\[
\mathbb{P}(P \mid G_s) = \mathbb{P}(P \mid G_g, S_1) = 0.40 = 40\%\,.
\]
Assuming that prior offenders constitute only 1\% of the general population, the generic likelihood ratio is $\mathbb{P}(P \mid G_g) / \mathbb{P}(P) = 0.74 / 0.01 = 74$, which suggests that profiling evidence provides strong support for the hypothesis that someone with a prior conviction committed \emph{some} burglary. However, the specific likelihood ratio is $\mathbb{P}(P \mid G_s) / \mathbb{P}(P) = 0.40 / 0.01 = 40$, which is nearly half the generic value. More importantly, if we did not know that the crime occurred in context $S_1$, we would have no justified estimate of $\mathbb{P}(P \mid G_s)$ at all. For all we know, it could be anywhere from 40\% to 95\%, depending on which hidden partition the crime falls into. This uncertainty renders the specific likelihood ratio effectively undefined in the absence of context-specific information.

\bibliographystyle{apalike}
\bibliography{biblioDvsP}

@article{mackor21,
  author       = {Anne Ruth Mackor},
  title        = {Different Ways of Being Naked. {A} Scenario Approach to the Naked
                  Statistical Evidence Problem},
  journal      = {{FLAP}},
  volume       = {8},
  number       = {9},
  pages        = {2407--2433},
  year         = {2021}
}

@article{enoch2021,
	author = {Enoch, David and Spectre, Levi},
	journal = {Synthese},
	number = {3},
	pages = {5687--5718},
	title = {Statistical resentment, or: what's wrong with acting, blaming, and believing on the basis of statistics alone},
		volume = {199},
	year = {2021}
	}

@article{Moss2023,
	author = {Sarah Moss},
	journal = {Oxford Studies in Epistemology},
	title = {Knowledge and Legal Proof},
        volume = {7},
	year = {2023}
}

@article{Gardiner2019,
	author = {Georgi Gardiner},
	doi = {10.1111/papa.12149},
	journal = {Philosophy and Public Affairs},
	number = {3},
	pages = {288--318},
	publisher = {Wiley},
	title = {The Reasonable and the Relevant: Legal Standards of Proof},
	volume = {47},
	year = {2019}
}

@article{dahlman2020,
author = {Christian Dahlman},
title ={Naked statistical evidence and incentives for lawful conduct},
journal = {The International Journal of Evidence \& Proof},
volume = {24},
number = {2},
pages = {162-179},
year = {2020},
doi = {10.1177/1365712720913333}
}

@book{Roush2006,
	author = {Roush, Sherrilyn},
	publisher = {Oxford University Press},
	title = {Tracking Truth: Knowledge, Evidence, and Science},
	year = {2006}
}

@article{pritchard2015,
	author = {Pritchard, Duncan},
	journal = {Metaphilosophy},
	number = {3},
	pages = {436-461},
	title = {Risk},
	volume = {46},
	year = {2015}
}

@book{moss2018,
	author = {Moss, Sarah},
	publisher = {Oxford University Press},
	title = {Probabilistic Knowledge},
	year = {2018}
}

@article{smith2017,
	author = {Smith, Martin},
	journal = {Mind},
	number = {508},
	pages = {1193--1218},
	title = {When Does Evidence Suffice for Conviction?},
	volume = {127},
	year = {2018}
}

@article{colyvan01,
	author = {Colyvan, Mark and Regan, Helen M. and Ferson, Scott},
	journal = {Journal of Political Philosophy},
	number = {2},
	pages = {168-181},
	title = {Is it a Crime to Belong to a Reference Class?},
	volume = {9},
	year = {2001}
}

@inbook{duff2021,
author = {Duff, Robin and Marshall, Sandra},
year = {2021},
pages = {77--91},
title = {Character, ‘Propensities’, and the (Mis)use of Statistics in Criminal Trials 1},
publisher= {In: \emph{The Social Epistemology of Legal Trials} edited by Hoskins, Z. and Robson, J.},
doi = {10.4324/9780429283123-5}
}

@article{Koehler1990Veridical-Verdi,
	author = {Koehler, Jonathan J. and Shaviro, Daniel N.},
	journal = {Cornell Law Review},
	pages = {247-279},
	title = {Veridical Verdicts: Increasing Verdict Accuracy Through the Use of Overtly Probabilistic Evidence and Methods},
	volume = {75},
	year = {1990}
}

@article{KoehelerBaseRateRelevant,
	author = {Koehler, Jonathan J.},
	journal = {Jurimetrics Journal},
	pages = {373-402},
	title = {When Do Courts Think Base Rate Statistics Are Relevant?},
	volume = {42},
	year = {2002}
}

@article{Banks2001Race-Bases-Susp,
	author = {Banks, Richard},
	journal = {UCLA Law Review},
	pages = {1075-1124},
	title = {Race-Bases Suspect Selection and Colorblind Equal Protection Doctrine and Discplusure},
	volume = {48},
	year = {2001}
}

@article{picinali16,
	author = {Picinali, Federico},
	journal = {Journal of Applied Philosophy},
	number = {1},
	pages = {69-87},
	title = {Base-rates of Negative Traits: instructions for Use in Criminal Trials},
	volume = {33},
	year = {2016}
}

@article{tillers05,
	author = {Tillers, Peter},
	journal = {Law, Probability and Risk},
	pages = {33-49},
	title = {If Wishes Were Horses: Discursive Comments on Attempts to Prevent Individuals from Being Unfairly Burdened by their Reference Classes},
	volume = {4},
	year = {2005}
}

@article{eggleston1980,
	author = {Eggleston, Richard},
	journal = {Criminal Law Review},
	pages = {678--688},
	title = {The Probability Debate},
	year = {1980}
}

@article{redmayne02,
	author = {Redmayne, Mike},
	journal = {Cambridge Law Journal},
	number = {3},
	pages = {684-714},
	title = {The relevance of Bad Character},
	volume = {61},
	year = {2002}
}

@article{pundik2017,
	author = {Pundik, Amit},
	journal = {Oxford Journal of Legal Studies},
	number = {1},
	pages = {189-216},
	title = {Freedom and Generalisation},
	volume = {37},
	year = {2017}
}

@article{Wasserman91,
	author = {Wasserman, David T.},
	journal = {Cardozo Law Review},
	pages = {935-976},
	title = {The Morality of Statistical Proof and the Risk of Mistaken Liability},
	volume = {13},
	year = {1991}
}

@book{Stein05,
	author = {Stein, Alex},
	publisher = {Oxford University Press},
	title = {Foundations of Evidence Law},
	year = {2005}
}

@book{jussim2012social,
  title={Social Perception and Social Reality: Why Accuracy Dominates Bias and Self-Fulfilling Prophecy},
  author={Jussim, Lee},
  year={2012},
  publisher={Oxford University Press},
  address={New York},
  isbn={9780195366600},
  doi={10.1093/acprof:oso/9780195366600.001.0001}
}

@article{thomsonIndEv1986,
    author = "Judy Thomson",
    title = "Liability and individualized evidence.",
    journal =  "Law and Contemporary Problems",
    volume = {49},
    number = {3},
    pages = {199-219},
    year = 1986 
}

@article{Lee-Stronach2023,
	author = {Chad Lee{-}Stronach},
	doi = {10.1111/nous.12486},
	journal = {No\^{u}s},
    volume = {58},
    number = {4},
    pages = {948--972},
	title = {Just Probabilities},
	year = {2024}
    
}

@book{schauer_profiles_2009,
	title = {Profiles, {Probabilities}, and {Stereotypes}},
	isbn = {978-0-674-04324-4},
	language = {en},
	publisher = {Harvard University Press},
	author = {Schauer, Frederick F.},
	month = jun,
	year = {2009},
}

@article{di_bello_profile_2020,
	title = {Profile {Evidence}, {Fairness}, and the {Risks} of {Mistaken} {Convictions}},
	volume = {130},
	doi = {10.1086/705764},
	number = {2},
	journal = {Ethics},
	author = {Di Bello, Marcello and O’Neil, Collin},
	year = {2020},
	pages = {147--178},
}

@article{enoch_statistical_2012,
	title = {Statistical {Evidence}, {Sensitivity}, and the {Legal} {Value} of {Knowledge}},
	volume = {40},
	number = {3},
	journal = {Philosophy \& Public Affairs},
	author = {Enoch, David and Spectre, Levi and Fisher, Talia},
	year = {2012},
	pages = {197--224}
}

\end{document}